\documentclass[twocolumn,prl,secnumarabic,
amsmath,amssymb,superscriptaddress,nofootinbib]
{revtex4-2}

\usepackage{graphicx} 
\usepackage{xcolor}
\usepackage{empheq}
\newcommand{\mL}{{\mathcal{L}}}

\newcommand{\mB}{{\mathcal{B}}}

\newcommand{\req}[1]{Eq.\,(\ref{#1})}

\begin{document}

\title{Comment on \lq\lq All-Loop Result for the Strong Magnetic Field Limit of the Heisenberg-Euler Effective Lagrangian\rq\rq} 

\author{Stefan Evans}
\affiliation{Department of Physics, The University of Arizona, Tucson, AZ 85721, USA}
\affiliation{Helmholtz-Zentrum Dresden-Rossendorf, Bautzner Landstraße 400, 01328 Dresden, Germany}
\author{Johann Rafelski}
\affiliation{Department of Physics, The University of Arizona, Tucson, AZ 85721, USA}

\begin{abstract}
We offer a 2nd look  at the recently claimed improvement of Euler-Heisenberg-Schwinger (EHS)  action [PRL \textbf{122} (2019) no.21, 211602 and post-publication correction]:  We demonstrate a difference to the claimed   concordance with the Schwinger-Dyson series starting at the two-loop  level.
\end{abstract}

\maketitle

In Ref.~\cite{Karbstein:2019wmj} a one-cut reducible loop summation extension of the Euler-Heisenberg-Schwinger (EHS)  action~\cite{Heisenberg:1935qt,Weisskopf:1996bu,Schwinger:1951nm} has been proposed, building on the recently developed two-loop  action~\cite{Gies:2016yaa}. 
We review these results and show that already at the two loop level the corresponding perturbative series is  not as claimed in Eqs.(8) and (9) of~\cite{Karbstein:2019wmj} a Schwinger-Dyson series. The primary concern of this discussion is the magnitude and sign of the two loop contribution; higher three and more loop contributions cannot be fully captured using summation techniques used in~\cite{Karbstein:2019wmj}. An extended resummation method is offered under separate cover~\cite{Evans:2023mxp}.

We recall the form of the renormalized (\lq r\rq ) one-loop EHS action in strong $\mB$ fields
\begin{align}
\label{EHSform}
\mL_1^{\rm r}=-\frac{\mB^2}2\Pi^{\rm r}_\mB\;,\quad
\Pi^{\rm r}_\mB= -\frac{e^2}{12\pi^2}\ln(\frac{e\mB}{m^2})
\;.
\end{align} 
The two-loop extension seen as Eq.\,(32) in~\cite{Gies:2016yaa} takes the form
\begin{align}
\label{2loop}
\mL_2^{\rm r} 
=&\;
\frac12\frac{\partial \mL_1^{\rm r}}{\partial F^{\mu\nu}}\frac{\partial \mL_1^{\rm r}}{\partial F_{\mu\nu}}
\;.
\end{align} 
We also draw attention to a post-publication correction seen in arXiv V4 release. An iterative procedure was subsequently proposed~\cite{Karbstein:2017gsb} for higher order loops, where the newly formed $\mL_2^{\rm r}$ takes the place of one of $\mL_1^{\rm r}$ on the RHS of \req{2loop}.

Using the strong $\mB$ field limit \req{EHSform},  in Eq.\,(8) of~\cite{Karbstein:2019wmj} a summation repeating the iteration  in~\cite{Karbstein:2017gsb} for a chain of one-cut reducible diagrams is presented  
\begin{align}
\label{assumedBEHS}
\mL_{\rm eff}^\mathrm{Karb}(\Pi^{\rm r}_\mB)  = 
-\frac{\mB^2}2+
\frac{\alpha(\Pi^{\rm r}_\mB)\mB^2}{6\pi}\ln(e\mB/m^2)
\;.
\end{align}    
Ref.~\cite{Karbstein:2017gsb} argues that a  field dependent  Schwinger-Dyson sum characterizes the  coupling constant $\alpha(\Pi^{\rm r}_\mB)$  prefactor to the one-loop EHS contribution, Eq.\,(9) in~\cite{Karbstein:2019wmj}: 
\begin{subequations}
\begin{align}
\label{running1st}
\alpha(\Pi^{\rm r}_\mB)=&\;
\frac{e^2}{4\pi}
\frac{1}{1+\Pi^{\rm r}_\mB}
\;.
\end{align}
This result was subsequently corrected in a post publication arXiv release without a formal erratum to read~\cite{KarbsteinV2:2019wmj}
\begin{align}
\label{running2nd}
\alpha(\Pi^{\rm r}_\mB)=&\;
\frac{e^2}{4\pi}\Big(1-\frac12 \frac{\Pi^{\rm r}_\mB}{1+\Pi^{\rm r}_\mB}\Big)
\;.
\end{align}
\end{subequations}

A better understanding of theoretical properties is achieved incorporating in the polarization the renormalization contribution with a logarithmic cutoff $\Lambda^2\gg e\mB$:  
\begin{align}
\label{EHSformBare}
\Pi^{\rm r}_\mB\to \Pi_\mB= \Pi_0+\Pi^{\rm r}_\mB,\quad 
\Pi_0= \frac{e^2}{12\pi^2}\ln(\frac{\Lambda^2}{m^2})
\;.
\end{align}  
$\Pi_\mB\propto -\ln e\mB/\Lambda^2$  is positive. The resulting $\alpha(\Pi_\mB)$ based on~\req{running1st}  satisfies the expectation that in the ultraviolet limit $\Pi_0\to \infty$ the coupling decreases continuously to zero. However, this is not the case for~\req{running2nd}  with the limit of  effective coupling reduced asymptotically by a factor two.

We return to consider the proposed  Lagrangian~\req{assumedBEHS} which includes the Maxwell contribution using the regularized (not renormalized) polarization function. Applying \req{running1st} to~\req{assumedBEHS}, we can cast the result into a simple analytical form allowing an easily checked perturbative expansion
\begin{subequations}
\begin{align}
\label{assumedBEHS1st}
\!\!\!
\mL_{\rm eff}^\mathrm{Karb}(\Pi_\mB) 
=&\;
-\frac{\mB^2}2\Big(1+
\frac{\Pi_\mB}{1+\Pi_\mB}\Big)
=-\frac{\mB^2}2\frac{1+2\Pi_\mB}{1+\Pi_\mB}
\\ \nonumber
=&\;
-\frac{\mB^2}2\Big(1+\Pi_\mB-\Pi_\mB^2+\Pi_\mB^3-{\mathcal{O}}(\Pi_\mB^4)\Big)
\;.
\end{align}  
Similarly using~\req{running2nd} we obtain another simple analytical form allowing another perturbative expansion
\begin{align}
\label{assumedBEHS2nd}
\mL_{\rm eff}^\mathrm{Karb}(\Pi_\mB) 
=&\;
-\frac{\mB^2}2\Big(1+\Pi_\mB
\Big(1-
\frac12\frac{\Pi_\mB}{1+\Pi_\mB}\Big)
\Big)
\\ \nonumber =&\;-\frac{\mB^2}2\frac{1+2\Pi_0+\Pi_0^2/2}{1+\Pi_0}
\\ \nonumber
 =&\;
 -\frac{\mB^2}2\Big(1+\Pi_\mB-\frac12\Pi_\mB^2+\frac12\Pi_\mB^3-{\mathcal{O}}(\Pi_\mB^4)\Big)
\;.
\end{align}  
\end{subequations}

For comparison, recall the usual Schwinger-Dyson series  
\begin{align}
\label{exampleSD}
\frac{1}{1-\Pi_0}
=1+\Pi_0+\Pi_0^2+\Pi_0^3+{\mathcal{O}}(\Pi_0^4)
\;.
\end{align}
A sign difference is evident in both published and corrected versions at the two-loop  order in $\Pi_0^2$:  While~\req{assumedBEHS1st} disagrees with the Dyson series by factor $-1$, the post-publication arXiv corrected~\req{assumedBEHS2nd} disagrees by factor $-1/2$.

In summary, we have shown that Ref.~\cite{Karbstein:2019wmj} proposes a loop summation which exhibits a sign (and magnitude considering the  post publication arXiv corrections) incompatibility with the
Schwinger-Dyson series at two loop level. We deal with other related matters under separate cover in~\cite{Evans:2023mxp}, where also a full Weisskopf-style evaluation of the EHS results is obtained.


\appendix

\end{document}